\documentclass[aps,prd,twoside,twocolumn,nofootinbib,showpacs,floatfix]{revtex4-1}
\usepackage{amssymb}
\usepackage{amsmath,amssymb}
\usepackage{graphicx,bm}
\usepackage{slashed}
\usepackage{times}
\usepackage{epstopdf}
\usepackage{ulem} 
\usepackage[usenames]{color}
\usepackage{float}
\usepackage{subfigure}
\usepackage{subfigure}
\usepackage{rotating}
\usepackage{color}
\usepackage{multirow}
\usepackage{dcolumn}
\usepackage{overpic}
\usepackage{booktabs}
\usepackage{makecell}
\usepackage{diagbox}

\renewcommand\sout{\bgroup \color{red} \ULdepth=-.5ex \ULset}

\usepackage[colorlinks, citecolor=blue,anchorcolor=red,menucolor=red,linkcolor=blue,filecolor=red,runcolor=red,urlcolor=blue,frenchlinks=red]{hyperref}
\begin{document}
\title{Systematics of the heavy flavor hadronic molecules}
\author{Kan Chen$^{1}$}
\author{Rui Chen$^{2}$}
\author{Lu Meng$^{3}$}
\author{Bo Wang$^{4,5}$}\email{wangbo@hbu.edu.cn}
\author{Shi-Lin Zhu$^{1}$}\email{zhusl@pku.edu.cn}

\affiliation{
$^1$School of Physics and Center of High Energy Physics, Peking University, Beijing 100871, China\\
$^2$Key Laboratory of Low-Dimensional Quantum Structures and Quantum Control of Ministry of Education, Department of Physics and Synergetic Innovation Center for Quantum Effects and Applications, Hunan Normal University, Changsha 410081, China\\
$^3$Institut f\"ur Theoretische Physik II, Ruhr-Universit\"at Bochum, D-44780 Bochum, Germany\\
$^4$School of Physical Science and Technology, Hebei University, Baoding 071002, China\\
$^5$Key Laboratory of High-precision Computation and Application of
Quantum Field Theory of Hebei Province, Baoding 071002, China }

\begin{abstract}
With a quark level interaction, we give a unified description of the
loosely bound molecular systems composed of the heavy flavor hadrons
$(\bar{D},\bar{D}^*)$, $(\Lambda_c, \Sigma_c, \Sigma_c^*)$, and
$(\Xi_c, \Xi_c^\prime,\Xi_c^*)$. Using the $P_c$ states as inputs to
fix the interaction strength of light quark-quark pairs, we
reproduce the observed $P_{cs}$ and $T_{cc}^+$ states and predict
another narrow $T_{cc}^{\prime+}$ state with quantum numbers
$[D^*D^*]_{J=1}^{I=0}$. If we require a satisfactory description of
the $T_{cc}^+$ and $P_c$ states simultaneously, our framework
prefers the assignments of the $P_{c}(4440)$ and $P_{c}(4457)$ as
the $[\Sigma_c\bar{D}^*]_{J=1/2}^{I=1/2}$ and
$[\Sigma_c\bar{D}^*]_{J=3/2}^{I=1/2}$ states, respectively. We
propose the isospin criterion to explain naturally why the
experimentally observed $T_{cc}$, $P_c$, and $P_{cs}$ molecular
candidates prefer the lowest isospin numbers. We also predict the
loosely bound states for the bottom di-hadrons.
\end{abstract}

\maketitle

\vspace{2cm}

\section{Introduction}\label{sec1}
The conventional mesons ($q\bar{q}$) and baryons ($qqq$) have been
extensively discussed at the birth of quark model
\cite{GellMann:1964nj,Zweig:19xx,Jaffe:1976ig}, and have become the
main part of hadron spectrum
nowadays~\cite{ParticleDataGroup:2020ssz}. Besides conventional
mesons and baryons, quantum chromodynamics (QCD) also allows the
existence of hadrons with more complicated configurations, such as
$qq\bar{q}\bar{q}$, $\bar{q}qqqq$, $qqqqqq$
($qqq\bar{q}\bar{q}\bar{q}$), etc.. In the past decades, many $XYZ$
states have been observed
~\cite{Chen:2016qju,Liu:2019zoy,Liu:2013waa,Guo:2017jvc,Hosaka:2016pey,Lebed:2016hpi,Esposito:2016noz,Brambilla:2019esw}
since the discovery of $X(3872)$ in 2003~\cite{Belle:2003nnu}. Some
of these states are below the thresholds of di-hadrons from several
to several tens MeVs. The molecular explanations were widely
proposed to understand their underlying structures.

In 2015, the LHCb Collaboration reported two structures $P_c(4380)$
and $P_c(4450)$ in the $J/\psi p$ mass spectrum~\cite{LHCb:2015yax}.
Their masses are consistent with the predictions of the hidden-charm
pentaquarks \cite{Wu:2010jy,Yang:2011wz,Wang:2011rga}. In 2019, the
LHCb Collaboration updated their analyses with larger data samples
and found that the $P_{c}(4450)$ consists of two $P_c$ states, i.e.,
$P_{c}(4440)$ and $P_{c}(4457)$~\cite{LHCb:2019kea}. Besides, they
further reported another near threshold hidden-charm pentaquark
$P_c(4312)$. These important results from LHCb provide strong
evidences for the existence of the hidden-charm molecular
pentaquarks
\cite{Chen:2019asm,Liu:2019tjn,He:2019ify,Xiao:2019aya,Meng:2019ilv,Yamaguchi:2019seo,PavonValderrama:2019nbk,Chen:2019bip,Burns:2019iih,Du:2019pij,Wang:2019ato}.
Thus, it is desirable to see whether there exist the hadronic
molecules in other heavy flavor di-hadron systems.

If enlarging the flavor symmetry group to SU(3), one may expect that
the $\Xi_c\bar{D}^{(*)}$, $\Xi^\prime_c\bar{D}^{(*)}$, and
$\Xi_c^*\bar{D}^{(*)}$ systems may also form molecular bound states.
The $P_{cs}$ pentaquarks were investigated in Refs.
\cite{Wu:2010jy,Chen:2016ryt,Santopinto:2016pkp,Shen:2019evi,Xiao:2019gjd,Wang:2019nvm}
and the most promising production channel $\Xi_b^-\rightarrow J/\psi
\Lambda K$ was suggested in Refs.
\cite{Santopinto:2016pkp,Chen:2015sxa}. Later, the LHCb reported the
evidence of $P_{cs}(4459)$ in the $J/\psi\Lambda$ invariant mass
spectrum~\cite{LHCb:2020jpq}, which agrees very well with the
prediction from the chiral effective field theory in Ref.
\cite{Wang:2019nvm}. However, this state still needs further
confirmation due to limited data samples at
present~\cite{LHCb:2020jpq}. Very probably, the $P_c$ and $P_{cs}$
pentaquarks share a very similar binding mechanism.

Very recently, the LHCb reported a very narrow structure
$T_{cc}^+(3875)$ in the $D^0D^0\pi^+$
spectrum~\cite{LHCb:2021auc,LHCb:2021vvq}. Its mass is slightly
below the $D^{*+}D^0$ threshold about $300$ keV. This signal tends
to confirm the predictions of the $DD^*$ molecular state with
quantum numbers $I(J^P)=0(1^+)$~\cite{Li:2012ss}. The
doubly heavy tetraquark states have been extensively studied and
debated in the literatures
\cite{Carlson:1987hh,Silvestre-Brac:1993zem,Semay:1994ht,Pepin:1996id,Gelman:2002wf,Vijande:2003ki,Janc:2004qn,Cui:2006mp,Navarra:2007yw,Vijande:2007rf,Ebert:2007rn,Lee:2009rt,Yang:2009zzp,Du:2012wp,Feng:2013kea,Ikeda:2013vwa,Molina:2010tx,Carames:2011zz,Sakai:2017avl,Luo:2017eub,Karliner:2017qjm,Eichten:2017ffp,Wang:2017uld,Cheung:2017tnt,Park:2018wjk,Li:2012ss,Xu:2017tsr}.
One can refer to Ref. \cite{Liu:2019zoy} for a review of the
$QQ\bar{q}\bar{q}$ system. This inspiring discovery also stimulated
a series of theoretical studies
\cite{Li:2021zbw,Weng:2021hje,Agaev:2021vur,Chen:2021vhg,Dai:2021wxi,Feijoo:2021ppq,Ling:2021bir,Meng:2021jnw,Wu:2021kbu,Yan:2021wdl,Hu:2021gdg,Jin:2021cxj,Yang:2021zhe,Ren:2021dsi,Chen:2021tnn,Azizi:2021aib,Fleming:2021wmk,Dong:2021bvy}.

The minimal valance quark components for the $P_c$, $P_{cs}$, and
$T_{cc}^+$ states are $c\bar{c}uud$, $c\bar{c}uds$, and
$cc\bar{u}\bar{d}$, respectively. Such states can not be
accommodated within the conventional quark model and thus give us a
golden platform to study the structures and dynamics of the
multiquark states.

We propose the following picture to understand the above heavy
flavor di-hadron systems. The heavy quark ($c$ or $b$) behaves like
a static color triplet source in each hadron. In the heavy quark
limit, its velocity $v$ does not change with time. The
non-relativistic property of heavy quarks is stimulative to
stabilize a molecular system
\cite{Lee:2011rka,Li:2012bt,Li:2014gra}. On the other hand, if the
particular combinations of the light quark components within the two
heavy hadrons can coincidentally provide the attractive force, then
the residual strong interaction that mainly comes from their light
degrees of freedom (d.o.f) may render this system to be bound. This
picture is similar to a hydrogen molecule in QED, where the two
electrons are shared by the two protons and the residual
electromagnetic force binds this system. Therefore, the question of
which heavy flavor di-hadron can form a bound state becomes which
kind of light quark combinations in the di-hadron system can provide
the enough attractive force.

In this work, we adopt the quark level interaction to address the
above question. We relate the heavy flavor di-hadron effective
potentials to their matrix elements in the flavor and spin spaces of
their light d.o.f, so that we can qualitatively determine which
heavy flavor di-hadron is possible to form a bound state.

This paper is organized as follows. In Sec. \ref{sec2}, we introduce
our theoretical framework. Then we present our numerical results and
discussions in Sec. \ref{sec3}. In Sec. \ref{sec4}, we conclude this
work with a short summary.

\section{Theoretical framework}\label{sec2}

We consider the two-body systems that are the combinations of the
ground-state hadrons $(\bar{D},\bar{D}^*)$, $(\Lambda_c, \Sigma_c,
\Sigma_c^*)$, and $(\Xi_c, \Xi_c^\prime,\Xi_c^*)$. We list the
physical allowed heavy flavor meson-meson, meson-baryon, and
baryon-baryon systems $[H_1H_2]_{J}^I$ in Table \ref{systems}, where
$H_1$ and $H_2$ denote the considered heavy flavor hadrons, while
$J$ and $I$ are the total angular momentum and total isospin of the
di-hadron system, respectively. If $H_1$ and $H_2$ are the general
identical particles (e.g., $\bar{D}\bar{D}$ and
$\bar{D}^*\bar{D}^*$), then the quantum numbers of the
$[H_1H_2]_{J}^I$ (fermion or boson) systems must satisfy the
following selection rule
\begin{eqnarray}\label{eq:srule}
L+S_{\text{tot}}+I_{\text{tot}}+2i=\text{Even number},\label{QN}
\end{eqnarray}
where $L$ is the orbital angular momentum between $H_1$ and $H_2$
($L=0$ for the $s$-wave case in our calculations), while
$S_{\text{tot}}$ and $I_{\text{tot}}$ are the total spin and total
isospin of the general identical di-hadron system, respectively. $i$
denotes the isospin of $H_1$ or $H_2$, e.g., $i=1/2$ for the $D$
meson, and $i=1$, $\frac{1}{2}$ for the $\Sigma_c$ and $\Xi_c$
baryons, respectively. The $2s$ ($s$ is the spin of $H_1$ or $H_2$)
is omitted from Eq. (\ref{QN}) since it must be an odd or even
number for the two-body systems of the identical fermions or bosons,
respectively.

\begin{table}[!htbp]
\renewcommand\arraystretch{1.5}
\caption{The allowed heavy flavor di-hadron systems
($[H_1H_2]^{I}_{J}$) that are considered in this work, where the
quantum numbers of the general identical systems are constrained by
Eq.~\eqref{eq:srule}.\label{systems}} \setlength{\tabcolsep}{2.1mm}
{
\begin{tabular}{clllccccccccccccccc}
\toprule[0.8pt]
\multirow{2}{*}{Meson-meson}&$[\bar{D}\bar{D}]^{1}_{0}$&$[\bar{D}\bar{D}^*]^{0,1}_{1}$&$[\bar{D}^*\bar{D}^*]^{1}_{0,2}$\\
&$[\bar{D}^*\bar{D}^*]^{0}_{1}$\\
\multirow{4}{*}{Baryon-meson}&$[\Lambda_c\bar{D}]^{\frac{1}{2}}_{\frac{1}{2}}$&$[\Lambda_c\bar{D}^*]_{\frac{1}{2},\frac{3}{2}}^{\frac{1}{2}}$&$[\Sigma_c\bar{D}]^{\frac{1}{2},\frac{3}{2}}_{\frac{1}{2}}$\\
&$[\Sigma_cD^*]_{\frac{1}{2},\frac{3}{2}}^{\frac{1}{2},\frac{3}{2}}$&$[\Sigma_c^*\bar{D}]_{\frac{3}{2}}^{\frac{1}{2},\frac{3}{2}}$&$[\Sigma^*_c\bar{D}^*]_{\frac{1}{2},\frac{3}{2},\frac{5}{2}}^{\frac{1}{2},\frac{3}{2}}$\\
&$[\Xi_c\bar{D}]_{\frac{1}{2}}^{0,1}$&$[\Xi_c\bar{D}^*]_{\frac{1}{2},\frac{3}{2}}^{0,1}$&$[\Xi_c^\prime\bar{D}]_{\frac{1}{2}}^{0,1}$\\
&$[\Xi_c^\prime\bar{D}^*]_{\frac{1}{2},\frac{3}{2}}^{0,1}$&$[\Xi_c^{*}\bar{D}]_{\frac{3}{2}}^{0,1}$&$[\Xi_c^*\bar{D}^*]_{\frac{1}{2},\frac{3}{2},\frac{5}{2}}^{0,1}$\\
\multirow{6}{*}{Baryon-baryon}&$[\Lambda_c\Lambda_c]^0_0$&$[\Lambda_c\Sigma_c]^{1}_{0,1}$&$[\Sigma_c\Sigma_c]^{0,2}_{0}$\\
&$[\Sigma_c\Sigma_c]^{1}_{1}$&$[\Lambda_c\Sigma_c^*]^1_{1,2}$&$[\Sigma_c\Sigma_c^*]_{1,2}^{0,1,2}$\\
&$[\Sigma_c^*\Sigma_c^*]_{1,3}^{1}$&$[\Sigma_c^*\Sigma_c^*]_{0,2}^{0,2}$&$[\Xi_c\Xi_c]_{0}^{1}$\\
&$[\Xi_c\Xi_c]_1^0$&$[\Xi_c\Xi_c^\prime]_{0,1}^{0,1}$
&$[\Xi_c\Xi_c^*]_{1,2}^{0,1}$\\
&$[\Xi_c^\prime\Xi_c^\prime]_{0}^{1}$&$[\Xi_c^\prime\Xi_c^\prime]_1^0$&$[\Xi_c^\prime\Xi_c^*]_{1,2}^{0,1}$\\
&$[\Xi_c^*\Xi_c^*]^0_{1,3}$&$[\Xi_c^*\Xi_c^*]^1_{0,2}$\\
\bottomrule[0.8pt]
\end{tabular}}
\end{table}

As discussed in the introduction, we assume that the interactions in
a heavy flavor di-hadron systems are mainly from the interactions of
their light quark components. We neglect the corrections from the
heavy d.o.f and study the residual strong interactions induced from
their light d.o.f. Then the heavy flavor meson-meson, meson-baryon,
and baryon-baryon can be studied simultaneously in the same
formalism by checking the interactions with all possible quantum
numbers.

We focus on the two-body $s$-wave interactions among the considered
ground heavy flavor mesons/baryons. One can formulate the
corresponding quark level Lagrangians as
\cite{Meng:2019nzy,Wang:2019nvm,Wang:2020dhf}
\begin{eqnarray}
\mathcal{L}=g_s\bar{q}\mathcal{S}q+g_a\bar{q}\gamma_\mu\gamma^5\mathcal{A}^\mu
q,\label{lag}
\end{eqnarray}
where $q=(u,d,s)$, $g_s$ and $g_a$ are two independent coupling
constants. They encode the nonperturbative dynamics between light
quarks of two color singlet hadrons and can be determined from the
experimental data.

From Eq. (\ref{lag}), we can see that the systems listed in Table
\ref{systems} can only couple to the isospin triplet and isospin
singlet fields. The systems that can couple to the strange isospin
doublet fields are not considered in the present work. Then the
fictitious scalar field $\mathcal{S}$ and axial-vector field
$\mathcal{A}^\mu$ reduce to the form
\begin{eqnarray}
\mathcal{S}&=&\mathcal{S}_3\lambda^i+\mathcal{S}_1\lambda^8,\label{scalar}\\
\mathcal{A}^\mu&=&\mathcal{A}_3^\mu\lambda^i+\mathcal{A}^\mu_1\lambda^8,\label{axial}
\end{eqnarray}
where $\lambda^i$ ($i=1,2,3$) and $\lambda^8$ are the generators of SU(3) group. $\mathcal{S}_3$ ($\mathcal{A}_3^\mu$) and $\mathcal{S}_1$ ($\mathcal{A}_1^\mu$) denote the isospin triplet and isospin singlet fields, respectively. 

The effective potential of light quark-quark interactions can be
deduced from Eq. (\ref{lag}), and we have
\begin{eqnarray}
V_{qq}&=&\tilde{g}_s\left(\lambda_1^8\lambda_2^8+\lambda_1^i\lambda_2^i\right)+\tilde{g}_a\left(\lambda_1^8\lambda_2^8+\lambda_1^i\lambda_2^i\right)\bm{\sigma}_1\cdot\bm{\sigma}_2.\nonumber\\
\label{Vqq}
\end{eqnarray}
where the effective potential $V_{qq}$ is reduced to the local form
when we integrate out the exchanged spurions (which is analogous to
the resonance saturation model \cite{Epelbaum:2001fm}). The
redefined coupling constants are $\tilde{g}_s=
g_s^2/m_{\mathcal{S}}^2$ and $\tilde{g}_a=g_a^2/m_{\mathcal{A}}^2$.
Then the heavy flavor di-hadron effective potential from the
interactions of their light quark components can be written as
\begin{eqnarray}
V_{[H_1H_2]_J^I}=\left\langle
[H_1H_2]_J^{I}\left|V_{qq}\right|[H_1H_2]_J^I\right\rangle,\label{VHH}
\end{eqnarray}
where $|[H_1H_2]_J^I\rangle$ denotes the quark-level spin-flavor
wave function of $H_1H_2$ system with total isospin $I$ and total
angular momentum $J$, which is the direct product of spin and flavor
wave functions
\begin{eqnarray}
\left|[H_1H_2]_J^I\right\rangle&=&\sum_{m_{I_1},m_{I_2}}C_{I_1,m_{I_1};I_2,m_{I_2}}^{I,I_z} \phi^{H_{1f}}_{I_1,m_{I_1}}\phi^{H_{2f}}_{I_2,m_{I_2}}\nonumber\\
&&\otimes\sum_{m_{l_1},m_{l_2}}C_{l_1,m_{l_1};l_2,m_{l_2}}^{l,l_z}\phi^{H_{1s}}_{l_1,m_{l_1}}\phi^{H_{2s}}_{l_2,m_{l_2}}.
\end{eqnarray}
The $C_{I_1,m_{I_1};I_2,m_{I_2}}^{I,I_z}$ and
$C_{l_1,m_{l_1};l_2,m_{l_2}}^{l,l_z}$ are the Clebsch-Gordan (CG)
coefficients. ($\phi^{H_{1f}}_{I_1,m_{I_1}}$,
$\phi^{H_{1s}}_{l_1,m_{l_1}}$), ($\phi^{H_{2f}}_{I_2,m_{I_2}}$,
$\phi^{H_{2s}}_{l_2,m_{l_2}}$) are the quark-level (flavor, spin)
wave functions for the $H_1$ and $H_2$ hadrons, respectively. $l$
denotes the total light spin of this two-body system.

From Eqs. \eqref{Vqq} and \eqref{VHH}, we can see that the residual
strong interaction of a specific $[H_1H_2]_J^I$ system can be
divided into four parts, i.e., the scalar type
($\lambda^8\lambda^8$), the isospin related type
($\lambda^i\lambda^i$), the spin related type
($\lambda^8\lambda^8\bm{\sigma}_1\cdot\bm{\sigma}_2$) and the
isospin-spin related type
$(\lambda^i\lambda^i\bm{\sigma}_1\cdot\bm{\sigma}_2)$ interactions.
In Table \ref{Mop}, we present the matrix elements of these four
types of operators for the considered di-hadron systems in Table
\ref{systems}.

\begin{table*}[htbp]
\scriptsize
\renewcommand\arraystretch{1.5}
\caption{The matrix elements of the operators $\mathcal{O}_1$
($\lambda^8_1\lambda^8_2$), $\mathcal{O}_2$
($\lambda^i_1\lambda^i_2$), $\mathcal{O}_3$
($\lambda^8_1\lambda^8_2\bm{\sigma}_1\cdot\bm{\sigma}_2$), and
$\mathcal{O}_4$
($\lambda^i_1\lambda^i_2\bm{\sigma}_1\cdot\bm{\sigma}_2$) for the
considered heavy flavor hadron-hadron systems ($[H_1H_2]_J^I$)
listed in Table \ref{systems}. \label{Mop}}
\setlength{\tabcolsep}{0.4mm}{
\begin{tabular}{|l|cccc|l|cccc|ccccccccc}
\toprule[0.4pt] \hline
System&$\mathcal{O}_1$&$\mathcal{O}_2$&$\mathcal{O}_3$&$\mathcal{O}_4$&System&$\mathcal{O}_1$&$\mathcal{O}_2$&$\mathcal{O}_3$&$\mathcal{O}_4$\\
\hline
$[\bar{D}\bar{D}]_0^1$&$\frac{1}{3}$&$1$&$0$&$0$&$[\bar{D}\bar{D}^*]^{0,1}_1$&$\frac{1}{3},\frac{1}{3}$&$-3,1$&$0,0$&$0,0$\\
$[\bar{D}^*\bar{D}^*]_{0,2}^1$&$\frac{1}{3},\frac{1}{3}$&$1,1$&$-\frac{2}{3},\frac{1}{3}$&$-2,1$&$[\bar{D}^*\bar{D}^*]_1^0$&$\frac{1}{3}$&$-3$&$-\frac{1}{3}$&$3$\\
\hline
$[\Lambda_c\bar{D}]^{\frac{1}{2}}_{\frac{1}{2}}$&$\frac{2}{3}$&$0$&$0$&$0$&$[\Lambda_c\bar{D}^*]^{\frac{1}{2}}_{\frac{1}{2},\frac{3}{2}}$&$\frac{2}{3},\frac{2}{3}$&$0,0$&$0,0$&$0,0$\\
$[\Sigma_c\bar{D}]_{\frac{1}{2}}^{\frac{1}{2},\frac{3}{2}}$&$\frac{2}{3},\frac{2}{3}$&$-4,2$&$0,0$&$0,0$&$[\Sigma_c\bar{D}^*]_{\frac{1}{2},\frac{3}{2}}^{\frac{1}{2}}$&$\frac{2}{3},\frac{2}{3}$&$-4,-4$&$-\frac{8}{9},\frac{4}{9}$&$\frac{16}{3},-\frac{8}{3}$\\
$[\Sigma_c\bar{D}^*]_{\frac{1}{2},\frac{3}{2}}^{\frac{3}{2}}$&$\frac{2}{3},\frac{2}{3}$&$2,2$&$-\frac{8}{9},\frac{4}{9}$&$-\frac{8}{3},\frac{4}{3}$&$[\Sigma_c^*\bar{D}]_{\frac{3}{2}}^{\frac{1}{2},\frac{3}{2}}$&$\frac{2}{3},\frac{2}{3}$&$-4,2$&$0,0$&$0,0$\\

$[\Sigma_c^*\bar{D}^*]_{\frac{1}{2},\frac{3}{2},\frac{5}{2}}^\frac{1}{2}$&$\frac{2}{3},\frac{2}{3},\frac{2}{3}$&$-4,-4,-4$&$-\frac{10}{9},-\frac{4}{9},\frac{2}{3}$&$\frac{20}{3},\frac{8}{3},-4$&$[\Sigma_c^*\bar{D}^*]_{\frac{1}{2},\frac{3}{2},\frac{5}{2}}^{\frac{3}{2}}$&$\frac{2}{3},\frac{2}{3},\frac{2}{3}$&$2,2,2$&$-\frac{10}{9},-\frac{4}{9},\frac{2}{3}$&$-\frac{10}{3},-\frac{4}{3},2$\\
\hline
$[\Xi_c\bar{D}]_{\frac{1}{2}}^{0,1}$&$-\frac{1}{3},-\frac{1}{3}$&$-3,1$&$0,0$&$0,0$&$[\Xi_c\bar{D}^*]^0_{\frac{1}{2},\frac{3}{2}}$&$-\frac{1}{3},-\frac{1}{3}$&$-3,-3$&$0,0$&$0,0$\\

$[\Xi_c\bar{D}^*]_{\frac{1}{2},\frac{3}{2}}^1$&$-\frac{1}{3},-\frac{1}{3}$&$1,1$&$0,0$&$0,0$&$[\Xi_c^\prime\bar{D}]_{\frac{1}{2}}^{0,1}$&$-\frac{1}{3},-\frac{1}{3}$&$-3,1$&$0,0$&$0,0$\\

$[\Xi_c^{\prime}\bar{D}^*]_{\frac{1}{2},\frac{3}{2}}^0$&$-\frac{1}{3},-\frac{1}{3}$&$-3,-3$&$\frac{4}{9},-\frac{2}{9}$&$4,-2$&$[\Xi_c^\prime\bar{D}^*]_{\frac{1}{2},\frac{3}{2}}^1$&$-\frac{1}{3},-\frac{1}{3}$&$1,1$&$\frac{4}{9},-\frac{2}{9}$&-$\frac{4}{3},\frac{2}{3}$\\

$[\Xi_c^*\bar{D}]_{\frac{3}{2}}^{0,1}$&$-\frac{1}{3},-\frac{1}{3}$&$-3,1$&$0,0$&$0,0$&$[\Xi_c^*\bar{D}^*]^0_{\frac{1}{2},\frac{3}{2},\frac{5}{2}}$&$-\frac{1}{3},-\frac{1}{3},-\frac{1}{3}$&$-3,-3,-3$&$\frac{5}{9},\frac{2}{9},-\frac{1}{3}$&$5,2,-3$\\

$[\Xi_c^*\bar{D}^*]_{\frac{1}{2},\frac{3}{2},\frac{5}{2}}^1$&$-\frac{1}{3},-\frac{1}{3},-\frac{1}{3}$&$1,1,1$&$\frac{5}{9},\frac{2}{9},-\frac{1}{3}$&$-\frac{5}{3},-\frac{2}{3},1$&&&&&\\

\hline
$[\Lambda_c\Lambda_c]_0^0$&$\frac{4}{3}$&$0$&$0$&$0$&$[\Lambda_c\Sigma_c]_{0,1}^1$&$\frac{4}{3},\frac{4}{3}$&$0,0$&$0,0$&$0,0$\\

$[\Sigma_c\Sigma_c]_0^{0,2}$&$\frac{4}{3},\frac{4}{3}$&$-8,4$&$-\frac{16}{9},-\frac{16}{9}$&$\frac{32}{3},-\frac{16}{3}$&$[\Sigma_c\Sigma_c]_1^{1}$&$\frac{4}{3}$&$-4$&$\frac{16}{27}$&$-\frac{16}{9}$\\

$[\Lambda_c\Sigma_c^*]_{1,2}^1$&$\frac{4}{3},\frac{4}{3}$&$0,0$&$0,0$&$0,0$&$[\Sigma_c\Sigma_c^*]^0_{1,2}$&$\frac{4}{3},\frac{4}{3}$&$-8,-8$&$-\frac{40}{27},\frac{8}{9}$&$\frac{80}{9},-\frac{16}{3}$\\

$[\Sigma_c\Sigma_c^*]^1_{1,2}$&$\frac{4}{3},\frac{4}{3}$&$-4,-4$&$-\frac{40}{27},\frac{8}{9}$&$\frac{40}{9},-\frac{8}{3}$&$[\Sigma_c\Sigma_c^*]^2_{1,2}$&$\frac{4}{3},\frac{4}{3}$&$4,4$&$-\frac{40}{27},\frac{8}{9}$&$-\frac{40}{9},\frac{8}{3}$\\

$[\Sigma_c^*\Sigma_c^*]^1_{1,3}$&$\frac{4}{3},\frac{4}{3}$&$-4,-4$&$-\frac{44}{27},\frac{4}{3}$&$\frac{44}{9},-4$&$[\Sigma_c^*\Sigma_c^*]^0_{0,2}$&$\frac{4}{3},\frac{4}{3}$&$-8,-8$&$-\frac{20}{9},-\frac{4}{9}$&$\frac{40}{3},\frac{8}{3}$\\

$[\Sigma_c^*\Sigma_c^*]^2_{0,2}$&$\frac{4}{3},\frac{4}{3}$&$4,4$&$-\frac{20}{9},-\frac{4}{9}$&$-\frac{20}{3},-\frac{4}{3}$&&&&&\\

\hline
$[\Xi_c\Xi_c]_0^1$&$\frac{1}{3}$&$1$&$0$&$0$&$[\Xi_c\Xi_c]_1^0$&$\frac{1}{3}$&$-3$&$0$&$0$\\

$[\Xi_c\Xi_c^\prime]_{0,1}^0$&$\frac{1}{3},\frac{1}{3}$&$-3,-3$&$0,0$&$0,0$&$[\Xi_c\Xi_c^\prime]_{0,1}^1$&$\frac{1}{3},\frac{1}{3}$&$1,1$&$0,0$&$0,0$\\

$[\Xi_c\Xi_c^*]_{1,2}^0$&$\frac{1}{3},\frac{1}{3}$&$-3,-3$&$0,0$&$0,0$&$[\Xi_c\Xi_c^*]_{1,2}^1$&$\frac{1}{3},\frac{1}{3}$&$1,1$&$0,0$&$0,0$\\

$[\Xi_c^\prime\Xi_c^\prime]_{0}^1$&$\frac{1}{3}$&$1$&$-\frac{4}{9}$&$-\frac{4}{3}$&$[\Xi_c^\prime\Xi_c^\prime]_1^0$&$\frac{1}{3}$&$-3$&$\frac{4}{27}$&$-\frac{4}{3}$\\

$[\Xi_c^\prime\Xi_c^*]_{1,2}^0$&$\frac{1}{3},\frac{1}{3}$&$-3,-3$&$-\frac{10}{27},\frac{2}{9}$&$\frac{10}{3},-2$&$[\Xi_c^\prime\Xi_c^*]_{1,2}^1$&$\frac{1}{3},\frac{1}{3}$&$1,1$&$-\frac{10}{27},\frac{2}{9}$&$-\frac{10}{9},\frac{2}{3}$\\

$[\Xi_c^*\Xi_c^*]_{1,3}^0$&$\frac{1}{3},\frac{1}{3}$&$-3,-3$&$-\frac{11}{27},\frac{1}{3}$&$\frac{11}{3},-3$&$[\Xi_c^*\Xi_c^*]_{0,2}^1$&$\frac{1}{3},\frac{1}{3}$&$1,1$&$-\frac{5}{9},-\frac{1}{9}$&$-\frac{5}{3},-\frac{1}{3}$\\
\bottomrule[0.8pt]
\end{tabular}
}
\end{table*}
After we obtain the effective potential of the $[H_1H_2]_J^I$
system, we need to check whether this system can form a bound state.
This can be achieved by solving the following Lippmann-Schwinger
equation (LSE),
\begin{eqnarray}
T(p^\prime,p)=V(p^\prime,p)+\int\frac{d^3q}{\left(2\pi\right)^3}\frac{V\left(p^\prime,q\right)T\left(q,p\right)}{E-\frac{
q^2}{2m_\mu}+i\epsilon},
\end{eqnarray}
where $m_\mu$ is the reduced mass of the $H_1$ and $H_2$. $p$ and
$p^\prime$ represent the momentum of the initial and final states in
the center of mass frame, respectively.

Here, we introduce a hard regulator to exclude the contributions
from higher momenta \cite{Meng:2021jnw,Meng:2021kmi}
\begin{eqnarray}
V(p,p^\prime)=V_{[H_1H_2]_J^I}\Theta\left(\Lambda-p\right)\Theta\left(\Lambda-p^\prime\right),
\end{eqnarray}
where $\Theta$ is the step function. The amplitude $T(p^\prime,p)$
is a function of $p^\prime$, $p$, and binding energy $E$ with a
separable form
\begin{eqnarray}
 T\left(p^\prime,p\right)=\beta(E)\Theta(\Lambda-p^\prime)\Theta(\Lambda-p).
\end{eqnarray}
Then the LSE can be reduced to an algebraic equation
\begin{eqnarray}
\beta(E)=\frac{V_{[H_1H_2]_J^I}}{1-V_{[H_1H_2]_J^I}G},\label{T}
\end{eqnarray}
with
\begin{eqnarray}
G=\frac{m_\mu}{\pi^2}\left[-\Lambda+k \tan^{-1}\left(\frac{\Lambda}{k}\right)\right], k=\sqrt{-2m_\mu E}.\nonumber\\
\end{eqnarray}
We can search for the pole position of Eq. (\ref{T}) to obtain the
binding energy of the $[H_1H_2]_J^I$ system.

\section{Numerical results}\label{sec3}
\subsection{The results of the $P_c$, $P_{cs}$, and $T_{cc}$ states}
We first use the masses of the $P_c(4312)$, $P_c(4440)$, and $P_c(4457)$
in Ref. \cite{LHCb:2019kea} to fix the parameters in our model.
In our previous work \cite{Wang:2019ato}, we suggested that
the $P_c(4312)$, $P_{c}(4440)$, and $P_c(4457)$ have the
assignments $[\Sigma_c\bar{D}]_{1/2}^{1/2}$, $[\Sigma_c\bar{D}^*]_{1/2}^{1/2}$,
and $[\Sigma_c\bar{D}^*]_{3/2}^{1/2}$, respectively. With the matrix
elements listed in Table \ref{Mop}, we can easily read out the
effective potentials for these three $P_c$ states,
\begin{eqnarray}
V_{P_{c}(4312)}&=&-\frac{10}{3}\tilde{g}_s,\\
V_{P_{c}(4440)}&=&-\frac{10}{3}\tilde{g}_s+\frac{40\tilde{g}_a}{9},\\
V_{P_{c}(4457)}&=&-\frac{10}{3}\tilde{g}_s-\frac{20\tilde{g}_a}{9}.
\end{eqnarray}

There exist three undetermined parameters in Eq. (\ref{T}), the
light quark-quark coupling constants $\tilde{g}_s$, $\tilde{g}_a$,
and the momentum cutoff $\Lambda$. We use the experimental mass of
the $P_c$ states \cite{LHCb:2019kea} to precisely extract these
three parameters. The solutions are $\tilde{g}_s=11.739$ GeV$^{-2}$,
$\tilde{g}_a=-2.860$ GeV$^{-2}$, and $\Lambda=0.409$ GeV. In our
convention, a positive (negative) $V_{[H_1H_2]_J^I}$ means a(n)
repulsive (attractive) interaction. Once we determine the signs of
the $\tilde{g}_s$ and $\tilde{g}_a$, we can directly find out
whether the considered systems have repulsive or attractive forces
from the values in Table \ref{Mop}.

From the point of view of the potential model, the spin-spin
interaction is suppressed by a factor of
$1/(m_{\Sigma_c^{(*)}}m_{\bar{D}^{(*)}})$, which is roughly
consistent with our obtained ratio
$|\tilde{g}_a|/|\tilde{g}_s|\approx 0.24$.

The cutoff $\Lambda$ is smaller than the masses of the ground scalar
or axial-vector mesons, which are regarded as the hard scales and
integrated out in the effective field theory. In principle, there
may exist contributions from the pion-exchange interactions.
Although the two ground heavy meson/baryons can easily couple to the
pion field via the $p$-wave interactions, our calculations based on
the chiral effective field theory
\cite{Chen:2021htr,Meng:2019ilv,Wang:2019ato,Wang:2019nvm,Wang:2020dhf}
showed that the magnitude of the one-pion-exchange (OPE) interaction
is comparable to that of the next-to-leading order two-pion-exchange
(TPE) interaction. The OPE and TPE have considerable corrections to
the binding energies of the bound states, but they are not the main
driving force of the formation of the bound states. In this work, we
do not include the pion exchange dynamics.

In Table \ref{BE}, we list the masses of the experimentally observed
molecular candidates and the results from our model. The center
values of the $P_c(4312)$, $P_{c}(4440)$, and $P_c(4457)$ are used
as inputs to determine the values of $\tilde{g}_s$, $\tilde{g}_a$,
and $\Lambda$. We also predict a $[\Sigma_c^*\bar{D}]^{1/2}_{3/2}$
molecular state with the mass 4376.2 MeV, which may correspond to
the observed $P_{c}(4380)$ state~\cite{LHCb:2015yax}. The bound
state $[\Sigma_c^*\bar{D}]^{1/2}_{3/2}$ is also obtained in
different models
\cite{Chen:2019asm,Liu:2019tjn,Xiao:2019aya,Yamaguchi:2019seo,PavonValderrama:2019nbk,Chen:2019bip,Du:2019pij,Wang:2019ato}.

\begin{table}[!htbp]
\renewcommand\arraystretch{1.5}
\caption{The experimental data
\cite{LHCb:2015yax,LHCb:2019kea,LHCb:2020jpq,LHCb:2021auc,LHCb:2021vvq}
and our results of the masses and binding energies (BE) for the
$T_{cc}^+$, $P_{c}(4312)$, $P_{c}(4380)$, $P_{c}(4440)$,
$P_{c}(4457)$, and $P_{cs}(4459)$. We adopt the isospin averaged
masses for the single-charm mesons and baryons
\cite{ParticleDataGroup:2020ssz}. The listed values are all in units
of MeV. \label{BE}} \setlength{\tabcolsep}{0.35mm}{
\begin{tabular}{c|cc|cccccccccccccccc}
\toprule[0.5pt] \hline
&Mass (Expt.)& BE (Expt.)& Mass (Our) &BE (Our)\\
\hline
$T_{cc}(3875)^+$&$3874.8$&$-1.0$&$3874.5$&$-1.8$\\
$P_{c}(4312)^+$&$4311.9\pm0.7^{+6.8}_{-0.6}$&$-8.9$ (input)&$4311.9$&$-8.9$\\
$P_{c}(4380)^+$&$4380\pm8\pm29$&$-6.2$&$4376.2$&$-9.1$\\
$P_{c}(4440)^+$&$4440.3\pm1.3^{+4.1}_{-4.7}$&$-21.8$ (input)&$4440.2$&$-21.8$\\
$P_{c}(4457)^+$&$4457.3\pm0.6^{+4.1}_{-1.7}$&$-4.8$ (input)&$4457.3$&$-4.8$\\
$P_{cs}(4459)^0$&$4458.8\pm2.9^{+4.7}_{-1.1}$&$-19.7$&$4468.1$&$-10.0$\\
\bottomrule[0.8pt]
\end{tabular}}
\end{table}

We further adopt our picture to study the recently observed
$T_{cc}^+$ state. We assign the $T_{cc}^+$ as the $[DD^*]_1^0$
molecular state and calculate its mass with the same parameters
extracted from the $P_c$ states. As shown in Table \ref{BE}, our
approach gives a rather good description of the mass of $T_{cc}^+$.
This nice agreement indicates that neglecting the corrections from
the heavy degrees of freedom is a fairly good approximation in this
case. The heavy flavor meson-meson and baryon-baryon systems share
the same binding mechanism that is dominated by their light degrees
of freedom.

The $P_{cs}(4459)$ is close to the threshold of $\Xi_c \bar{D}^*$,
which can be assigned as a $[\Xi_c \bar{D}^*]^0_{1/2}$ or $[\Xi_c
\bar{D}^*]_{3/2}^{0}$ molecular state \cite{Wang:2019nvm}. The spin
of the light diquark in the $\Xi_c$ baryon is 0, so the $\Xi_c
\bar{D}^*$ system has the vanishing spin-spin interaction from its
light d.o.f. The $[\Xi_c \bar{D}^*]^0_{1/2}$ and $[\Xi_c
\bar{D}^*]_{3/2}^{0}$ states are degenerate in our formalism, as can
be seen from Table \ref{Mop}. The inclusion of the spin-spin
interaction from heavy degrees of freedom or pion-exchange shall
distinguish these two states, which is beyond the scope of the
present work. However, from a serious calculation within the
framework of chiral effective field theory \cite{Wang:2019nvm}, the
mass gaps induced from the spin-spin interactions are within several
MeVs. In this sense, our prediction is consistent with the observed
$P_{cs}(4459)$. Indeed, the LHCb collaboration also fitted the data
using two resonances with masses $4454.9\pm2.7$ MeV and
$4467.8\pm3.7$ MeV \cite{LHCb:2020jpq}. However, the limited data
samples cannot confirm or refute the two-peak hypothesis. An updated
analysis with more data samples is desired to clarify this issue.

We further swap the assignments of the $P_c(4440)$ and
$P_{c}(4457)$, and regard them as the
$[\Sigma_c\bar{D}^*]_{3/2}^{1/2}$ and
$[\Sigma_c\bar{D}^{*}]_{1/2}^{1/2}$ molecular states, respectively.
We can also find a set of solutions that can reproduce the masses of
the three $P_c$ states. However, the cutoff $\Lambda$ is at 1.763
GeV, which is far away from the scale of light scalar mesons.
Moreover, we can not reproduce the $T_{cc}^+$ state in this case.
Thus, we rule out this set of assignments for the $P_c(4440)$ and
$P_c(4457)$. In our framework, we can identify the quantum numbers
of the $P_{c}(4440)$ and $P_{c}(4457)$ states if we require a
satisfactory description of the $T_{cc}^+$ and $P_c$ states
simultaneously.

\subsection{$T_{cc}^\prime$ state and other heavy flavor molecular states}

In the previous section, we have shown that our framework gives a
nice description of the observed $T_{cc}$, $P_c$, and $P_{cs}$
states. In the following, we further adopt the fitted parameters
$\tilde{g}_s$, $\tilde{g}_a$, and $\Lambda$ to calculate the other
heavy flavor di-hadron systems listed in Table \ref{systems}. The
effective potentials for these systems can be easily read from Table
\ref{Mop}. Their calculated masses and binding energies are listed
in Table \ref{ccBE}.

\begin{table*}[!htbp]
\renewcommand\arraystretch{1.5}
\caption{The predicted masses and binding energies (BE) for the
charmed di-hadrons ($[H_1H_2]_J^{I}$) in Table \ref{systems}. We
adopt the isospin averaged masses for the single-charm hadrons
\cite{ParticleDataGroup:2020ssz}. The values are all in units of
MeV. \label{ccBE}} \setlength{\tabcolsep}{1.5mm}{
\begin{tabular}{c|cccccccccccccccccc}
\toprule[0.8pt]
System&$[\bar{D}^*\bar{D}^*]_{1}^0$
&$[\Sigma_c^*\bar{D}^*]_{\frac{1}{2}}^\frac{1}{2}$
&$[\Sigma_c^*\bar{D}^*]_{\frac{3}{2}}^{\frac{1}{2}}$
&$[\Sigma_c^*\bar{D}^*]_{\frac{5}{2}}^\frac{1}{2}$
&$[\Xi_c\bar{D}]_{\frac{1}{2}}^0$
&$[\Xi_c\bar{D}^*]_{\frac{3}{2}}^0$  &$[\Xi_c^\prime
\bar{D}]_{\frac{1}{2}}^0$  &$[\Xi_c^\prime
\bar{D}^*]_{\frac{1}{2}}^0$&$[\Xi_c^\prime
\bar{D}^*]^0_{\frac{3}{2}}$&$[\Xi_c^*\bar{D}]_{\frac{3}{2}}^0$&$[\Xi_c^*\bar{D}^*]_{\frac{1}{2}}^0$
\\
\hline
Mass  &$4009.7$&$4501.3$&$4510.1$&$4523.8$&$4327.7$&$4468.1$&$4436.7$&$4564.9$&$4582.1$&$4503.6$&$4628.5$\\
BE&$-7.4$&$-25.4$&$-15.9$&$-2.9$&$-8.9$&$-10.0$&$-9.4$&$-22.5$&$-5.2$&$-9.6$&$-26.0$\\
System&$[\Xi_c^*\bar{D}^*]_{\frac{3}{2}}^0$&
$[\Xi_c^*\bar{D}^*]_{\frac{5}{2}}^0$&$[\Sigma_c\Sigma_c]_0^0$&$[\Sigma_c\Sigma_c]^1_{1}$&$[\Sigma_c\Sigma_c^*]_1^0$&$[\Sigma_c\Sigma_c^*]_2^0$&$[\Sigma_c\Sigma_c^*]^1_1$&$[\Sigma_c\Sigma_c^*]^1_2$
&$[\Sigma_c^*\Sigma_c^*]_{1}^1$&$[\Sigma_c^*\Sigma_c^*]_3^1$&$[\Sigma_c^*\Sigma_c^*]_0^0$
\\
Mass&$4638.0$&$4651.3$&$4825.4$&$4903.9$&$4894.3$&$4931.9$&$4958.4$&$4969.3$&$5021.9$&$5035.1$&$4946.5$\\
BE&$-16.5$&$-3.2$&$-81.7$&$-3.2$&$-77.4$&$-39.8$&$-13.3$&$-2.4$&$-14.4$&$-1.2$&$-89.8$\\
System&$[\Sigma_c^*\Sigma_c^*]_2^0$&$[\Xi_c\Xi_c]^0_1$
&$[\Xi_c\Xi_c^\prime]_{0,1}^0$&$[\Xi_c\Xi_c^*]^0_{1,2}$&$[\Xi_c^\prime\Xi_c^\prime]_1^0$&$[\Xi_c^{\prime}\Xi_c^*]_1^0$&$[\Xi_c^\prime\Xi_c^*]_2^0$&$[\Xi_c^*\Xi_c^*]_1^0$&$[\Xi_c^*\Xi_c^*]_3^0$\\
Mass&$4996.1$&$4933.1$&$5042.1$&$5109.0$&$5153.6$&$5210.3$&$5221.7$&$5276.4$&$5290.3$\\
BE&$-40.1$&$-5.8$&$-6.2$&$-6.4$&$-4.0$&$-14.4$&$-3.0$&$-15.6$&$-1.7$\\
\bottomrule[0.8pt]
\end{tabular}
}
\end{table*}

We check all the physically allowed $D^{(*)}D^{(*)}$ systems and
find that there exists another $T_{cc}^{\prime+}$ state with
$[D^*D^*]_1^0$ assignment. The $T_{cc}^{\prime+}$ lies about 7 MeV
below the $D^*D^*$ threshold. This state has no hidden-charm strong
decay channels due to its $cc\bar{u}\bar{d}$ valance quark
component, thus should decay into $D^0D^0\pi^0\pi^+$ or
$D^0D^+\pi^0\pi^0$ final states. In addition, because of the small
phase space for $D^{*+}\rightarrow D^+\pi^0$ ($D^0\pi^+$) and
$D^{*0}\rightarrow D^0\pi^0$, if the $T_{cc}^{\prime+}$ does exist,
similar to the $T_{cc}^+$, the $T_{cc}^{\prime+}$ should also be a
narrow state in a loosely bound molecular picture. We suggest the
LHCb Collaboration to look for this state in the future.

The results for the charmed baryon-meson (baryon) systems are also
presented in Table \ref{ccBE}. The determined $\tilde{g}_s$ is a
positive value and  about $3$ times larger than the $\tilde{g}_a$.
From Eq. (\ref{Vqq}) we can see that for the lowest isospin
di-hadron systems, the isospin-isospin matrix elements are negative
and dominate the whole effective potentials of the di-hadron
systems. Thus, if a heavy flavor two-body system has a large
negative $\lambda^i_1\lambda_2^i$ eigenvalue, this two-body system
will have an attractive force and may form a bound state. As shown
in Table \ref{Mop} and \ref{ccBE}, this feature is universal for all
the studied heavy flavor meson-meson, meson-baryon, and
baryon-baryon systems. The $\lambda_1^i\lambda_2^i$ reduces to
$\bm{\tau}_1\cdot\bm{\tau}_2$ in the SU(2) case, and can be
calculated with
\begin{eqnarray}
\bm{\tau}_1\cdot\bm{\tau}_2=2\left[I(I+1)-I_1(I_1+1)-I_2(I_2+1)\right].\label{tautau}
\end{eqnarray}
As shown in Eq. (\ref{tautau}), the lowest total isospin generally
leads to a negative eigenvalue and corresponds to an attractive
force. Our formalism gives a very practical criterion to understand
why the currently observed $T_{cc}^+$, $P_c$, and $P_{cs}$ states
all prefer the lowest isospins.

\subsection{Implications for the bottom hadron-hadron systems}

In our calculations, we neglect the corrections from the heavy
quarks in the charmed two-body systems and obtain a good description
of the observed $T_{cc}^+$, $P_c$, and $P_{cs}$ states. If we adopt
the same approximation for the bottom di-hadrons, then the
($T_{cc}^+$, $P_{c}$, $P_{cs}$) and ($T_{bb}^-$, $P_{b}$, $P_{bs}$)
molecular states share the identical effective potentials from their
light d.o.f.

\begin{figure}[htbp]
    \centering
    \includegraphics[width=1.0\linewidth]{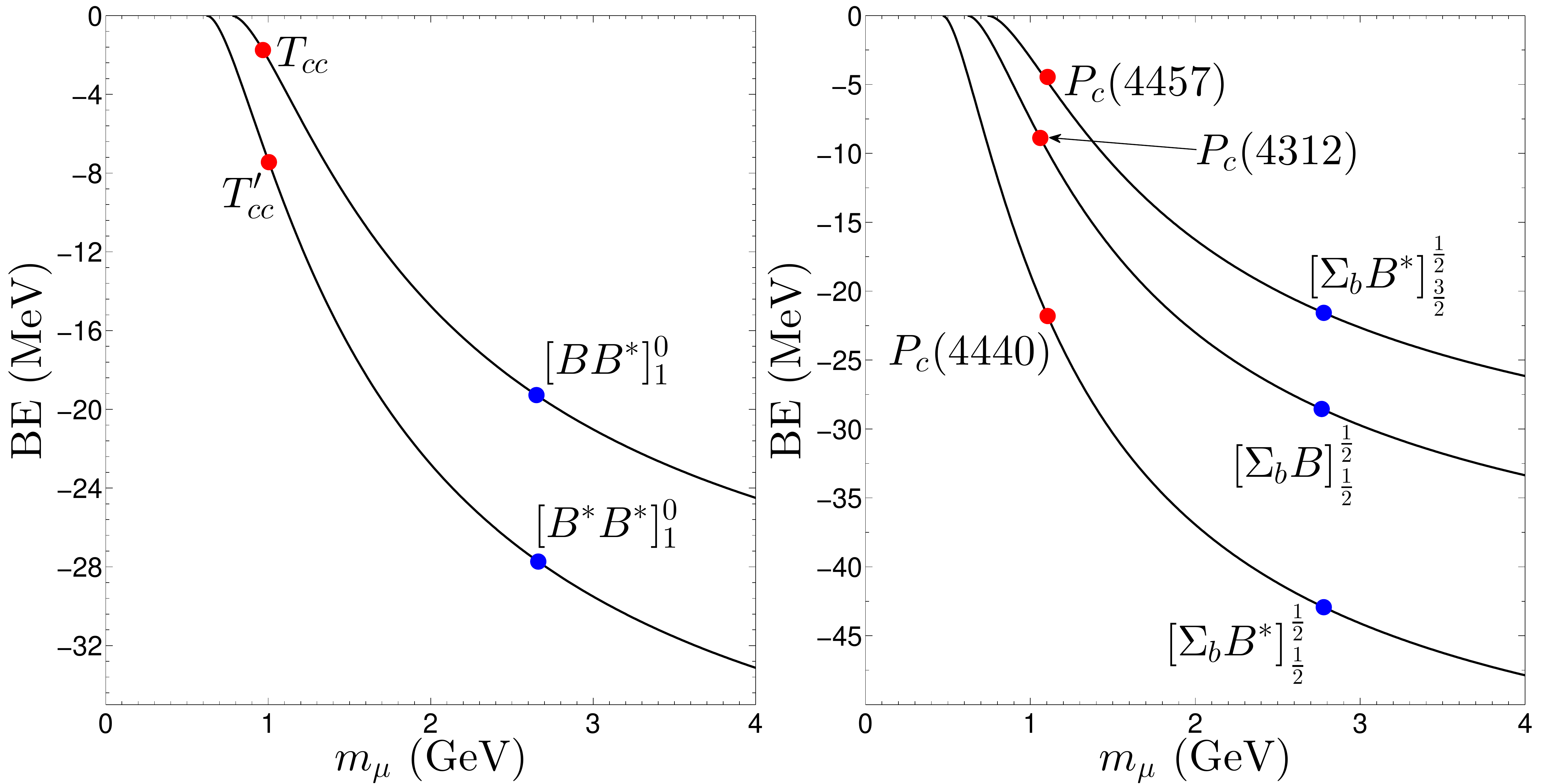}
    \caption{The variation of binding energies for the $T_{cc}^+$, $T_{cc}^{\prime+}$, $P_{c}(4312)$, $P_c(4440)$, and $P_{c}(4457)$ states as their reduced masses increase. At $m_Q=m_b$, we have their bottom partners $T_{bb}^-$, $T_{bb}^{\prime-}$, $[\Sigma_bB]_{1/2}^{1/2}$, $[\Sigma_bB^*]_{1/2}^{1/2}$, and $[\Sigma_bB^*]_{3/2}^{1/2}$, respectively.}
    \label{mctomb}
\end{figure}

In Fig. \ref{mctomb}, we present the variation of binding energies
for some typical molecular states as their corresponding reduced
masses gradually increase. In each system, there exists a critical
reduced mass at $E_{\text{BE}}=0$, from which the system starts to
form a bound state. Then the absolute values of binding energies
increase as their reduced masses increase. The increased rate
depends on the different types of light quark combinations in the
two-body heavy flavor systems. We mark the $T_{cc}^+$,
$T_{cc}^{\prime+}$, $P_{c}(4312)$, $P_{c}(4440)$, and $P_{c}(4457)$,
as well as their bottom partners in Fig. \ref{mctomb}. For the rest
of the considered bottom meson-meson, meson-baryon and baryon-baryon
systems, we list our predictions in Table \ref{bbBE}. As shown in
Fig. \ref{mctomb}, due to the large reduced masses of the bottom
di-hadron systems, if there exist bound states in the charm
di-hadrons, there should also exist the bottom partners with deeper
binding energies as well.

\begin{table*}[!htbp]
\renewcommand\arraystretch{1.5}
\caption{The predicted masses and binding energies (BE) for the
bottom di-hadrons ($[H_1H_2]_J^{I}$) in Table \ref{systems}. We
adopt the isospin averaged masses for the single-bottom hadrons
\cite{ParticleDataGroup:2020ssz}. The values are all in units of
MeV. \label{bbBE}} \setlength{\tabcolsep}{0.82mm}{
\begin{tabular}{c|cccccccccccccccccc}
\toprule[0.8pt]
System&$[BB^*]_1^0$&$[B^*B^*]_1^0$&$[\Sigma_bB]_{\frac{1}{2}}^{\frac{1}{2}}$&$[\Sigma_bB^*]_{\frac{1}{2}}^{\frac{1}{2}}$
&$[\Sigma_bB^*]_{\frac{3}{2}}^{\frac{1}{2}}$&$[\Sigma_b^*B]_{\frac{3}{2}}^{\frac{1}{2}}$&$[\Sigma_b^*B^*]_{\frac{1}{2}}^{\frac{1}{2}}$
&$[\Sigma_b^*B^*]_{\frac{3}{2}}^{\frac{1}{2}}$&$[\Sigma_b^*B^*]_{\frac{5}{2}}^{\frac{1}{2}}$&$[\Xi_bB]_{\frac{1}{2}}^0$
&$[\Xi_bB^*]^0_{\frac{1}{2},\frac{3}{2}}$&$[\Xi_b^\prime
B]_{\frac{1}{2}}^0$
\\
\hline
Mass&$10584.9$&$10621.7$&$11064.1$&$11094.9$&$11116.2$&$11083.5$&$11110.7$&$11121.5$&$11139.1$&$11048.0$&$11093.1$&$11181.2$\\
BE&$-19.3$&$-27.7$&$-28.5$&$-42.9$&$-21.6$&$-28.6$&$-46.6$&$-35.8$&$-18.1$&$-28.5$&$-28.6$&$-33.3$\\
System&$[\Xi_b^\prime B^*]^0_{\frac{1}{2}}$&$[\Xi_b^\prime
B^*]_{\frac{3}{2}}^0$&$[\Xi_b^*B]_{\frac{3}{2}}^0$&$[\Xi_b^*B^*]^0_{\frac{1}{2}}$&$[\Xi_b^*B^*]_{\frac{3}{2}}^0$
&$[\Xi_b^*B^*]_{\frac{5}{2}}^0$&$[\Sigma_b\Sigma_b]_0^0$&$[\Sigma_b\Sigma_b]_1^1$&$[\Sigma_b\Sigma_b^*]_1^0$&$[\Sigma_b\Sigma_b^*]_{2}^0$
&$[\Sigma_b\Sigma_b^*]_1^1$&$[\Sigma_b\Sigma_b^*]_2^1$
\\
Mass&$11216.6$&$11238.0$&$11206.1$&$11233.3$&$11244.1$&$11261.8$&$11523.2$&$11609.3$&$11547.5$&$11586.4$&$11615.6$&$11630.5$\\
BE&$-43.1$&$-21.7$&$-28.7$&$-46.7$&$-35.9$&$-18.3$&$-103.0$&$-16.9$&$-98.2$&$-59.3$&$-30.0$&$-15.1$\\
System&$[\Sigma_b^*\Sigma_b^*]_1^1$&$[\Sigma_b^*\Sigma_b^*]_3^1$&$[\Sigma_b^*\Sigma_b^*]_0^0$&$[\Sigma_b^*\Sigma_b^*]_2^0$&$[\Xi_b\Xi_b]_1^0$&$[\Xi_b\Xi_b^\prime]^0_{0,1}$&$[\Xi_b\Xi_b^*]_{1,2}^0$
&$[\Xi_b^\prime\Xi_b^\prime]_1^0$&$[\Xi_b^\prime
\Xi_b^*]_1^0$&$[\Xi_c^\prime\Xi_c^*]_2^0$
&$[\Xi_b^*\Xi_b^*]^0_{1}$&$[\Xi_b^*\Xi_b^*]_3^0$
\\
Mass&$11634.1$&$11652.6$&$11554.7$&$11605.8$&$11573.4$&$11711.3$&$11731.6$&$11852.9$&$11860.0$&$11875.0$&$11879.4$&$11898.0$\\
BE&$-31.0$&$-12.4$&$-110.4$&$-59.3$&$-20.6$&$-20.7$&$-20.7$&$-17.2$&$-30.3$&$-15.4$&$-31.3$&$-12.7$\\
\bottomrule[0.8pt]
\end{tabular}
}
\end{table*}

\section{Summary}\label{sec4}

In this work, we use a quark level effective potential to give a
universal description of the heavy flavor hadronic molecules that
are composed of the ground $(\bar{D},\bar{D}^*)$, $(\Lambda_c,
\Sigma_c, \Sigma_c^*)$, and $(\Xi_c, \Xi_c^\prime,\Xi_c^*)$ hadrons.
Based on this quark-level effective Lagrangian, we neglect the
contributions from heavy quarks and relate the effective potentials
of di-hadrons to their flavor and spin interaction operators of
light degrees of freedom.

In our approach, we only introduce three parameters $\tilde{g}_s$,
$\tilde{g}_a$, and $\Lambda$. They can be well extracted from the
observed $P_{c}(4312)$, $P_c(4440)$, and $P_c(4457)$.  We exclude
the assignments of $P_{c}(4440)$ and $P_{c}(4457)$ as the
$[\Sigma_c\bar{D}^*]_{3/2}^{1/2}$ and
$[\Sigma_c\bar{D}^*]_{1/2}^{1/2}$ states, respectively, due to the
poor description of $T_{cc}^+$ in this case. Our results strongly
indicate a very similar binding mechanism between the heavy flavor
meson-meson and meson-baryon systems, i.e., they are bound
dominantly by the interactions of their light degrees of freedom. We
further generalize this similarity to the heavy flavor baryon-baryon
systems.

We predict another $T_{cc}^{\prime+}$ state with the assignment
$[D^\ast D^\ast]_{1}^0$. From our calculations, the $T_{cc}^+$ and
$T_{cc}^{\prime+}$ are the only two molecular states in the
$D^{(*)}D^{(*)}$ systems. We suggest the LHCb to look for this state
in the future. We also predict other possible heavy flavor hadronic
molecules in the charmed and bottom sectors (e.g., see Tables
\ref{ccBE} and \ref{bbBE}).

\section*{Acknowledgments}
K. Chen thanks Zi-Yang Lin for helpful discussions. B. Wang is
supported by the Youth Funds of Hebei Province (No. 042000521062)
and the Start-up Funds for Young Talents of Hebei University (No.
521100221021). This project is also supported by the National
Natural Science Foundation of China under Grants 11975033 and
12070131001.

\end{document}